\documentclass[aps,pre,amsmath,amssymb,twocolumn,superscriptaddress,showpacs]{revtex4}

\usepackage{graphicx}

\begin{document}
\title{Quantum-thermal annealing with cluster-flip algorithm}

\author{Satoshi Morita}
\affiliation{International School for Advanced Studies (SISSA),
Trieste 34151, Italy}

\author{Sei Suzuki}
\affiliation{Department of Physics and Mathematics, Aoyama Gakuin University,
Fuchinobe, Sagamihara 229-8558, Japan}

\author{Tota Nakamura}
\affiliation{College of Engineering, Shibaura Institute of Technology,
Minuma-ku, Saitama 330-8570, Japan}


\begin{abstract}
A quantum-thermal annealing method using a cluster-flip
algorithm is studied in the two-dimensional spin-glass model. 
The temperature ($T$) and the transverse field ($\Gamma$) are decreased
simultaneously with
the same rate along a linear path on the $T$-$\Gamma$ plane.
We found that the additional pulse of the transverse field
to the frozen local spins
produces a good approximate solution with a low computational cost.
\end{abstract}

\pacs{02.70.Ss, 02.70.Uu}

\maketitle

An optimization problem frequently appears in many situations of science
and technology \cite{GareyJohnson}.
All known algorithms
need an exponentially-long time to find the exact solution of
non-deterministic polynomial problems.
The traveling salesman problem, the satisfiability problem, and the 
ground-state search problem of spin glasses are the typical examples. 
It is important to develop an efficient algorithm that provides an
approximate solution with a reasonable computational time.
A standard method for such problems is the classical simulated-annealing 
method (CA), which was proposed based on an analogy between optimization
problems and statistical mechanics \cite{KirkpatrickGV,Cerny}.
The idea of CA is to identify the cost function to be minimized with the
energy in a statistical-mechanical system.
We perform a numerical simulation to obtain the equilibrium state of
this system.
The temperature is artificially introduced into the system and gradually
decreased toward zero. 
The thermal fluctuations enable the system to escape from the local minima
of the cost function. At low temperatures, the system may stay with a high
probability in the optimal state.

Quantum annealing (QA) is a novel algorithm proposed as an alternative of
CA \cite{KadowakiNishimori, SantoroMTC, DasChakrabarti, SantoroTosatti}.
We use the quantum fluctuations instead of the thermal fluctuations. 
The quantum fluctuations are introduced by the additional energy term that is
non-commutative with the cost function of the optimization problem. 
In the Ising spin model, the transverse magnetic field ($\Gamma$)
serves as this additional term.
The additional term is initially set to take a large value so that the
initial state is disordered by the quantum fluctuations.
It is gradually decreased with time, and it finally vanishes. 
Only the cost function term remains in the final state.  
It is expected that QA can find an approximate solution more
efficiently than CA because of the quantum tunneling between the local minima.
Analytical \cite{MoritaN2006, MoritaN2007, Suzuki2008} and numerical
\cite{DasChakrabarti, SantoroTosatti, MartonakST} investigations 
support this scenario.

\begin{figure}
 \includegraphics[width=4cm,clip]{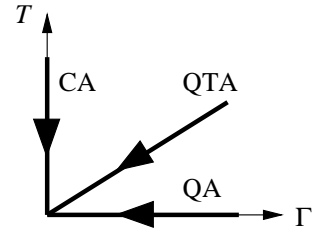}
\caption{Paths for three types of annealing
method in the plane of the temperature and the transverse field.
Note that, when the quantum Monte-Carlo method is applied to QA in
 practice, the QA path is lifted from the $\Gamma$-axis because a small
 but non-zero temperature is introduced. }
\label{fig:diagram}
\end{figure}

Although QA and CA have been studied a lot so far, 
hybrid methods of them have not attracted much attentions. 
Lee and Berne \cite{LeeBerne} 
applied a method which carries out QA and CA alternately 
to the protein folding problem. 
Battaglia \textit{et al}. \cite{Battaglia} examined a similar method 
to the satisfiability problem.
The present paper aims to investigate quantum-thermal annealing
(QTA) that carries out QA and CA \textit{simultaneously}.
In particular, we focus on the linear path
on the transverse field ($\Gamma$) v.s. temperature ($T$) 
plane where these two parameters are lowered with
keeping $\beta\Gamma$ constant (Fig. \ref{fig:diagram}),
where $\beta = 1/T$.
To investigate QTA numerically in large systems,
we used the path-integral Monte-Carlo method (PIMC) with the
cluster-flip algorithm (the cluster-flip PIMC) \cite{NakamuraIto,Nakamura}. 
There are
two reasons why the cluster-flip PIMC is suitable for QTA:
first, the cluster-flip PIMC can simulate any 
Ising-spin system in the transverse field without not only the error
due to Suzuki-Trotter mapping \cite{Trotter,Suzuki1971} 
but also the slowdown of relaxation known as the Wiesler freezing
\cite{Wiesler}.
Second, the computational cost is independent of the temperature and
the transverse field if keeping $\beta\Gamma$ constant.

The performance of QTA by the cluster-flip PIMC
(the cluster-QTA) is governed by the parameter $\beta\Gamma$.
The residual error is small but the computational cost becomes larger
when the value of $\beta\Gamma$ is large.
To solve this dilemma, we introduce {\it the local transverse-field
pulse process}.
The main result of the present paper is that 
the cluster-QTA with the local pulse process exhibits the best performance in
comparison among various annealing methods with regard to the residual
energy and the computational cost.

Let us consider the transverse field Ising model defined by the Hamiltonian,
\begin{equation}
 H = -\sum_{\langle ij\rangle}J_{ij}\sigma_i^z \sigma_j^z
  -\Gamma \sum_{i=1}^{N}\sigma_i^x,
\label{eq:H}
\end{equation}
where $\sigma_i^\alpha$ $(\alpha=x,y,z)$ are the Pauli matrices
of the spin $1/2$ operator at a site $i$. The first term
corresponds to the cost function of the optimization problem we want
to solve. 
The quenched random interaction $J_{ij}$ follows the Gaussian distribution
with the average zero and the variance one.  
We use, for the Monte-Carlo simulation,
a single configuration of random couplings on the
$100\times 100$ system with the periodic boundary condition. 
We confirmed that the results given below 
are consistent with similar simulations with other 
random configurations on $50\times 50$ and $30\times 30$ systems. 
The second term is the transverse field, 
which introduces the quantum fluctuation to the system.

In PIMC, the transverse field Ising model is mapped to the classical
Ising model with an additional imaginary-time dimension by the
Suzuki-Trotter formula \cite{LandauBinder,Trotter, Suzuki1971}. 
The partition function
at the inverse temperature $\beta$ is expressed as
\begin{eqnarray}
 Z_M &=& {\rm Tr}\ \exp\left[\sum_{m=1}^{M}\left(
  \frac{\beta}{M}\sum_{\langle ij\rangle}J_{ij}S_i^{(m)}S_j^{(m)} 
 \right.\right. \nonumber \\
 && \left.\left.
  +\gamma\sum_{i=1}^{N}S_i^{(m)}S_i^{(m+1)}
		      \right)\right],
  \label{eq:Z}
\end{eqnarray}
where $M$ is the length along the extra dimension (Trotter number) and
$S_i^{(m)}$ $(=\pm 1)$ denotes a classical Ising spin at a site $i$ on
the $m$th Trotter slice. The boundary condition along the imaginary-time
direction is periodic, {\it i.e.}, $S_i^{(m+M)}=S_i^{(m)}$. The
nearest-neighbor interaction between adjacent Trotter slices,
\begin{equation}
 \gamma=\frac{1}{2}\ln\left(\coth \frac{\beta \Gamma}{M}\right) ,
\end{equation}
is ferromagnetic.
The mapped classical system is equivalent to the
original quantum system in the limit of the infinite Trotter number.
As far as the Trotter number is finite, the error of
quantum-classical mapping appears.
To suppress the error at low temperature,
one has to choose a large Trotter number. 
When the Trotter number is large,
the simulation suffers from the Wiesler freezing \cite{Wiesler},
where the spin-flip probability becomes very small and the spin state 
freezes in a practical computational time.

The cluster-flip algorithm proposed in Refs. \cite{NakamuraIto, Nakamura}
solves the problems of the standard PIMC.
The idea of the present algorithm is to make an update
global in the imaginary-time direction and local in the real space.
We create clusters only in the imaginary-time direction in 
the same manner as the Swendsen-Wang method \cite{SwendsenWang}. 
Each cluster is updated according to the local molecular-field from the
neighboring sites. Since the correlated cluster in the imaginary-time direction
is flipped by one update trial, the Wiesler freezing does not
occur. Moreover, one can take the infinite limit of the
Trotter number by replacing dynamics of clusters themselves
to that of domain walls between clusters.
It follows that the error of the Suzuki-Trotter mapping
disappears.
We can identify the spin states of the mapped classical system
by the location of the domain walls and the spin state on the lowest
Trotter slice.
The computational cost of the cluster-flip PIMC is
proportional to $\beta\Gamma$, which is the average  number
of domain walls in the imaginary-time direction at one site.
This is contrasted with the standard single-flip PIMC where
the computational cost is governed by the Trotter number $M$.

In QTA, we employ the schedule of the transverse field given by
\begin{equation}
 \Gamma(t)=\Gamma_0(1- t/\tau),
  \label{eq:schedule}
\end{equation}
where $\Gamma_0$ is the initial value and $\tau$ is the annealing time. 
The Monte-Carlo step $t$ moves from $t=0$ to $t=\tau$. 
We decrease the transverse field and the temperature
simultaneously while keeping $\beta\Gamma$ constant.
Note that the convergence annealing schedule for QTA is
the inverse-logarighmic law of the temperature and the transverse field
\cite{Convergence}, which is slower than the power-law schedule for QA. 
However, the system surely converges to the
desired optimal state in the infinite time limit.
We also consider the standard annealing
schedule of QA for comparison, 
in which the transverse field is decreased as
Eq. (\ref{eq:schedule}) with the temperature
fixed at a sufficiently low value.

The initial cluster states in QTA and QA are made by giving the
position of domain walls by the Poisson process with the 
mean value $\beta\Gamma$. 
If the number of generated domain walls, $n_{\rm dw}$, is odd, 
one of the domain walls is removed to obey the periodic
boundary conditions along the imaginary-time direction.
Since the expectation value of $\sigma^x$ is $n_{\rm dw}/\beta
\Gamma$ in the infinite limit of the Trotter number, 
the initial state corresponds to the
all-up state along the $x$ axis.

\begin{figure}
 \includegraphics[width=8.5cm]{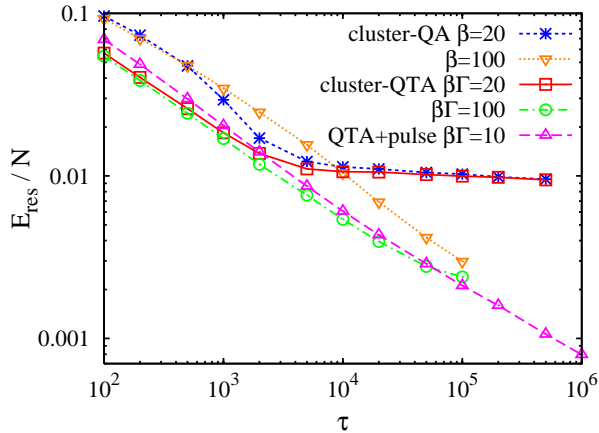}
 
 \caption{(Color online) Annealing time $\tau$ dependence of the minimum residual 
 energy per site for a $100\times 100$ two-dimensional Ising spin glass. 
 The cluster-QTA is performed with $\beta\Gamma = 20$ and $100$.
 Correspondingly, the cluster-QA is done with $\beta=20$ and $100$.
 The broken curve with upper triangles is 
 the data obtained by the cluster-QTA with the local pulse process
 performed with $\beta\Gamma=10$.
 The initial value of the transverse field 
 is chosen to be $\Gamma_0=1$ in all cases.
 Each annealing method is tried at least 100 times.}

 \label{fig:Eres}
\end{figure}

Figure \ref{fig:Eres} shows minimum residual energies of QTA and QA 
by the cluster-flip PIMC against the annealing time $\tau$. 
The minimum residual energy is defined by the difference between
the lowest energy in all the Trotter slices and the ground-state energy.
The ground-state energy was obtained by the spin-glass server
\cite{sgserver}.
The parameters $\beta\Gamma$ for QTA and $\beta$ for QA are set at
$\beta\Gamma=\beta=20$ or $\beta\Gamma=\beta=100$.
The initial value of the transverse field is chosen at $\Gamma_0 = 1$
in all cases. 
Every simulation is repeated at least 100 times.

\begin{figure}
 \includegraphics[width=8.5cm]{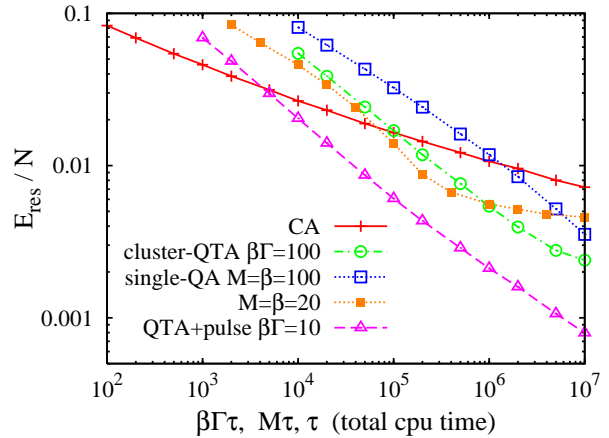}

 \caption{(Color online) Comparison of the minimum residual energies per site 
obtained by the
cluster-QTA, single-QA and CA. The abscissa is the computational cost
(total cpu time) which is defined by the annealing time $\tau$ 
multiplied by 
$\beta \Gamma$ for the cluster-QTA, $\tau$ multiplied by $M$ for the 
single-QA, 
and $\tau$ for CA. The initial temperature for CA is $T_0=3$ and the initial
 transverse field for the other methods is set $\Gamma_0=1$.
}

 \label{fig:shift}
\end{figure}

As common to QTA and QA, when $\beta\Gamma$ or $\beta$ is small,
the residual energy rapidly decays for small $\tau$,
while it is almost saturated for large $\tau$.
The change in the residual energy curve from a rapid
decay to a slow decay is understood as
a crossover from the intrinsic curve of QA to the curve of
QA with strong thermal disturbances \cite{PSARS}.
The cluster-QTA with $\beta\Gamma=100$ preserves the large decay-rate
longer than that with $\beta\Gamma=20$.
Since the thermal fluctuation is less effective when $\beta\Gamma$ is 
large, the rapid decay is attributed to QA at zero temperature. 
On the contrary, the thermal fluctuation begins to dominate the dynamics
as $\tau$ becomes large.
The system finds itself at the finite temperature after the long
Monte-Carlo steps.
This is because the rate of non-adiabatic transition is suppressed and
the excitation occurs due to very small but finite thermal fluctuations.
We observed that the similar crossover takes place also in the
QA by the single-flip PIMC (see data of $\beta=M=20$ in Fig.~\ref{fig:shift}), 
as it has been recognizable in Fig.~1 of Ref. \cite{MartonakST}.

For large $\tau$, after the crossover takes place,
the final state of the cluster-QTA is a classical state
in which no domain wall is present at all. 
Such a state loses the quantum fluctuations represented by $n_{\rm dw}/\beta\Gamma$.
Once a spin falls into a classical state 
which minimizes the cost function locally,
it can hardly escape from the local minimum.
Therefore, we introduce the local pulse process.

The local pulse process is applied to a real-space site when there
is no domain wall along the imaginary-time direction at this site.
What is done in this process is to re-initialize the cluster state, so
that it contains $\beta\Gamma$ domain walls on average as in the
disordered initial state.
The other sites are left untouched.
This process is physically interpreted as an application of the on-site
transverse-field pulse.
The pulse process destroys classical states by creating domain
walls and induces extra quantum fluctuations.
It helps the spin state to escape from the local minimum.
An increase in energy caused by this process is on the order of unity
because it is applied locally.
It is noted that every site always has at least two domain walls during
the whole annealing process.
However, this constraint does not matter for the spin-glass model we
have investigated.
Because of the global spin-flip symmetry the final spin state is
expected to contain both the ground state and its spin-flip state.
For the system without the spin-flip symmetry, we might need to stop the
pulse process at some time.

The broken curve connecting upward triangles in Fig.~\ref{fig:Eres} 
shows the performance of the cluster-QTA with the local pulse process.  
Even if we choose $\beta\Gamma=10$, the residual energy continues
to decay without the crossover.
We found that the parameter $\beta\Gamma=10$ is almost optimal for the
cluster-QTA with the pulse process.  
If $\beta\Gamma$ is larger than 10,
a larger computational cost is needed
to obtain the same residual energy. 
On the other hand, the smaller $\beta\Gamma$ causes insufficient
quantum fluctuations which raise the residual energy.
The additional cost of computation by the pulse process is only about
the factor of $0.2$ of the whole computational time.
It is because the pulse process is not always applied and the
re-initialization of the cluster state does not cost as much as the
update.

Figure \ref{fig:shift} shows the residual energy against the
computational cost.
We performed not only the cluster-QTA but also CA and
QA by the single-flip PIMC (single-QA) on the same sample.
In CA, we chose a completely random
spin configuration as an initial state and the temperature is 
decreased linearly from the initial value, $T_0=3$, to zero. 
As for the single-QA, 
the initial spin configuration is completely random in the
$(2+1)$-dimensional space.
The inverse temperature $\beta$ is fixed to the
Trotter number $M$ according to Ref. \cite{MartonakST}, 
and the initial value of the transverse field is set at $\Gamma_0=1$. 
These initial values for $T$ and $\Gamma$ are almost optimized 
so as to obtain the lowest residual energy.
The residual energy in the single-QA is obtained from the lowest
energy among all the Trotter slices. 
The computational cost is defined by $\tau$ multiplied by
$\beta\Gamma$ for the cluster-QTA, $\tau$ multiplied by $M$ for 
the single-QA, 
and $\tau$ for CA, which roughly correspond to the total cpu time.

Apparently the decay rate of CA (solid curve) is slower than others.
The single-QA and the cluster-QTA with and without the pulse process
have similar decay rates. 
However, the cluster-QTA with the pulse process achieves the highest
efficiency compared with other methods.
It is clearly observed both in Figs.~\ref{fig:Eres} and \ref{fig:shift}
that the crossover appears when a value of $\beta\Gamma$ or $\beta$ is
small.
We consider that it appears much later in case when $\beta\Gamma$ is
large.
The crossover prevents us from obtaining a lower residual energy.
We must choose a large value of $\beta\Gamma$ in order to avoid the
crossover.
However, we must pay much computational cost in turn.
The cluster-QTA with the local pulse process is the solution of
this dilemma.
We put the quantum fluctuations not to the whole system but only to the
spin that is frozen earlier.
This is the main reason why we can achieve the lower residual energy
very efficiently.

In conclusion, we investigated QTA implemented
by the cluster-flip PIMC
and revealed that the local transverse-field pulse improves the
performance of QTA. In practice, we showed that QTA with the pulse process 
gives the best performance with respect to the residual energy
and the computational cost, compared to CA, 
QA, and QTA without the pulse process.
Although we have considered only the two-dimensional Ising
spin-glass, it is noteworthy that QA by
the single-flip method was reported 
not to be better than CA in a satisfiability 
problem \cite{Battaglia}.
Whether QTA by the cluster-flip method
works well in such a harder optimization problem is the future problem.

\acknowledgments

The authors are grateful to G.~E.~Santoro and H.~Nishimori for helpful
discussions.
S.M. is supported by JSPS Postdoctoral Fellowships for
Research Abroad. 
S.S. acknowledges support by Grant-in-Aid for Scientific Research 
(Grant No. 20740225) from the Ministry of Education, Culture, Sports,
Science and Technology of Japan.

\end{document}